\documentclass[prl,superscriptaddress,reprint]{revtex4-1}

\usepackage{amsmath}
\usepackage{graphicx}
\usepackage{bm}

\usepackage[usenames,dvipsnames,x11names]{xcolor}
\usepackage{tikz}
\usetikzlibrary{arrows}

\newcommand{\be}{\begin{equation}}
\newcommand{\ee}{\end{equation}}
\newcommand{\bea}{\begin{eqnarray}}
\newcommand{\eea}{\end{eqnarray}}
\newcommand{\ket}[1]{\left\vert #1    \right\rangle }
\newcommand{\bra}[1]{\left\langle   #1  \right\vert}
\newcommand{\ave}[1]{\left\langle #1   \right\rangle }
\newcommand{\prj}[1]{\left\vert #1    \right\rangle \left\langle   #1  \right\vert}

\newcommand{\x}{^{\dagger}}
\newcommand{\w}{\omega}
\newcommand{\W}{\Omega}
\newcommand{\e}{\epsilon}
\newcommand{\da}{\downarrow}
\newcommand{\ua}{\uparrow}

\begin{document}

\title{Spin Coherent Manipulation in Josephson Weak Links}
\author{Javier Cerrillo}
\email{javier.cerrillo@upct.es}
\address{\'Area de F\'isica Aplicada, Universidad Polit\'ecnica de Cartagena, E-30202 Cartagena, Spain}
\address{Departamento de F\'isica Te\'orica de la Materia Condensada C-V, Universidad Aut\'onoma de Madrid, E-28049 Madrid, Spain}
\author{M.~Hays}
\affiliation{Department of Applied Physics, Yale University, New Haven, CT 06520, USA}
\author{V.~Fatemi}
\affiliation{Department of Applied Physics, Yale University, New Haven, CT 06520, USA}
\author{Alfredo Levy Yeyati}
\email{a.l.yeyati@uam.es}
\address{Departamento de F\'isica Te\'orica de la Materia Condensada C-V, Universidad Aut\'onoma de Madrid, E-28049 Madrid, Spain}
\address{Condensed Matter Physics Center (IFIMAC) and Instituto Nicol\'as Cabrera, Universidad Aut\'onoma de Madrid, E-28049 Madrid, Spain}

\begin{abstract}
Novel designs of Josephson weak links based on semiconducting nanowires combined with circuit QED techniques have enabled the resolution of their fine structure due to spin-orbit interactions, opening a path towards Andreev spin qubits. Nevertheless, direct manipulation of the spin within a given Andreev state is in general suppressed compared to inter-doublet manipulation in the absence of Zeeman effects. In addition, noisy spin-flip mechanisms limit any coherent manipulation protocol to spin post-selection. We propose a combination of a spin polarization protocol analogous to sideband cooling with stimulated Raman adiabatic passage specifically tailored for these systems. We show this approach is robust for a large range of design parameters, including the currently rather stringent coherence times.
\end{abstract}

\maketitle

While conventional superconducting qubits are becoming a mature technology 
\cite{Arute2019},
other radically new approaches involving superconducting nanostructures remain to be explored \cite{Karzig2017}.
In particular some qubit proposals rely on the Andreev bound states (ABS) which appear in highly 
transmissive few-channel Josephson weak links \cite{Desposito2001,Zazunov2003,Janvier2015,Chtchelkatchev2003,Padurariu2010,Park2017}. Within these proposals one can 
distinguish those operating when the ABS are populated with an even 
number of particles (the even sector), which are called Andreev level qubits (ALQ) \cite{Desposito2001,Zazunov2003,Janvier2015}; and the ones in which an extra unpaired quasiparticle is trapped in the weak link, which corresponds to the so-called Andreev spin qubits (ASQ) \cite{Chtchelkatchev2003,Padurariu2010,Park2017}. This type of sytems combines 
interesting aspects of microscopic degrees of freedom (such as their isolation from the environment) 
with the potential scalability of solid state devices. Another attractive aspect of ASQ is their inherently fermionic  
character and their connection to Majorana physics \cite{Prada2019}. In fact, under  
sufficiently large magnetic fields the proximitized nanowires used in  
ASQ experiments could host Majorana states at their edges \cite{Lutchyn2010,Oreg2010}.

\begin{figure}[hb!]
\centerline{
    \begin{tikzpicture}[
      level/.style={very thick},
      virtual/.style={ultra thin,densely dashed},
      transthb/.style={ultra thick,blue!50!black,<->,shorten >=2pt,shorten <=2pt,>=stealth},
      transb/.style={ultra thin,blue!50!black,<->,shorten >=2pt,shorten <=2pt,>=stealth},
      transbl/.style={ultra thin,<->,shorten >=2pt,shorten <=2pt,>=stealth},
      transthbl/.style={ultra thick,<->,shorten >=2pt,shorten <=2pt,>=stealth},
      transthr/.style={ultra thick,color=orange!50!red,<->,shorten >=2pt,shorten <=2pt,>=stealth},
    ]
    \node[] (russell) at (3.7cm,-1em)
    {\includegraphics[width=\columnwidth]{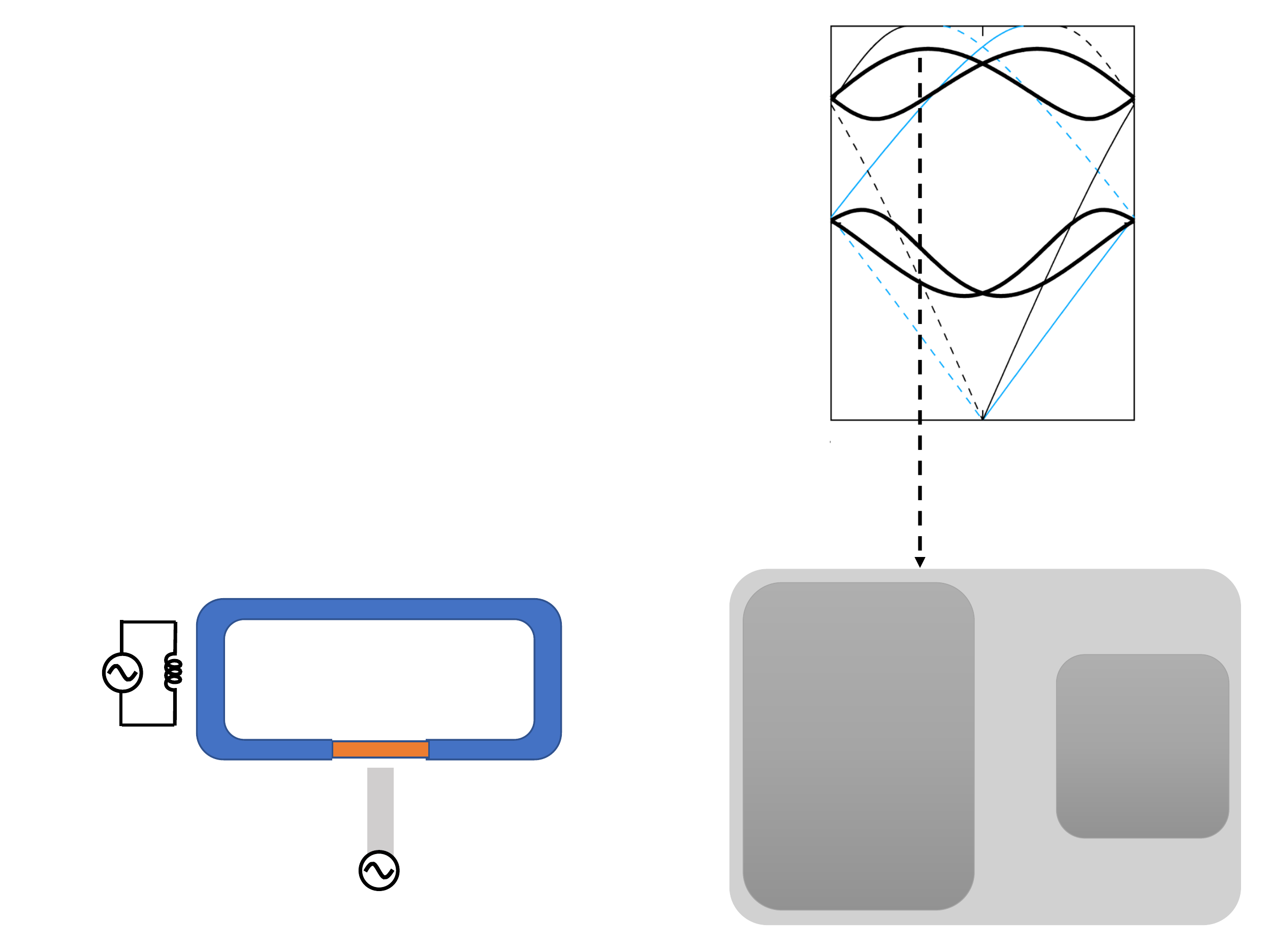}};
    \draw[level] (0cm,.5em) -- (1cm,.5em) node[midway,below] {$\ket{A}$};
    \draw[level] (2cm,2em) -- (3cm,2em) node[midway,below] {$\ket{B}$};
    \draw[level] (1cm,7em) -- (2cm,7em) node[midway,above] {$\ket{C}$};
    \draw[virtual] (1cm,5em) -- (2cm,5em) node[midway,above] {};
    \draw[transthb] (.5cm,.5em) -- (1.4cm,5em) node[midway,left] {$\W_A$};
    \draw[transthr] (2.5cm,2em) -- (1.6cm,5em) node[midway,right] {$\W_B$};
    \draw[transbl] (1.5cm,5em) -- (1.5cm,7em) node[midway,left] {$\Delta$};
    \node[draw=none,rectangle]  at (-.3cm,8em) {a)};
    \node[draw=none,rectangle]  at (-.3cm,-3em) {b)};
    \node[draw=none,rectangle]  at (3.8cm,8em) {c)};
    \node[draw=none,rectangle]  at (3.8cm,-3em) {d)};
    \node[draw=none,rectangle, rotate=90]  at (4.3cm,4em) {$\epsilon_{ks}/\Delta_S$};
    \node[draw=none,rectangle]  at (4.8cm,.4em) {0.0};
    \node[draw=none,rectangle]  at (4.8cm,4.4em) {0.5};
    \node[draw=none,rectangle]  at (4.8cm,8.4em) {1.0};
    \node[draw=none,rectangle]  at (6cm,-1.5em) {$\phi/\pi$};
    \node[draw=none,rectangle]  at (5.1cm,-0.5em) {0};
    \node[draw=none,rectangle]  at (6.1cm,-0.5em) {1.0};
    \node[draw=none,rectangle]  at (7.1cm,-0.5em) {2.0};
    \node[draw=none,rectangle]  at (2cm,-5.2em) {\Large$\Phi$};
    \node[draw=none,rectangle]  at (-.2cm,-5.2em) {$v_f$};
    \node[draw=none,rectangle]  at (1.5cm,-9.5em) {$v_g$};
    \draw[level] (5.8cm,-9em) -- (5cm,-9em) node[left] {$\ket{1}$};
    \draw[level] (5.8cm,-4.5em) -- (5cm,-4.5em) node[left] {$\ket{2}$};
    \draw[level] (6.7cm,-6.5em) -- (7.65cm,-6.5em) node[midway,above] {$\ket{\da}$};
    \draw[level] (6.7cm,-7em) -- (7.65cm,-7em) node[midway,below] {$\ket{\ua}$};
    \draw[transthb] (5.2cm,-9em) -- (5.2cm,-4.7em) node[midway,left] {$\W_g$};
    \draw[transthr] (5.4cm,-9em) -- (5.4cm,-5.5em) node[midway,right] {$\W_f$};
    \draw[transb] (5.95cm,-6em) -- (6.65cm,-6em) node[midway,above] {$\alpha\W_g$};
    \draw[transthbl] (5.95cm,-7em) -- (6.65cm,-7em) node[midway,below] {$g$};
    \end{tikzpicture}
}
\caption{a) Traditional STIRAP setup with a three-level system in $\Lambda$-configuration and a Raman coupling generated by two fields characterized by Rabi frequencies $\W_A$ and $\W_B$. b) Schematic diagram of the experimental setup: a superconducting loop (blue) threaded by a magnetic flux $\Phi$ is weakly linked by a semiconducting nanowire (orange). The weak link can be controlled by AC gate voltage $v_g$ and inductively by an AC flux voltage $v_f$. c) Level dispersion relation as a function of $\phi$ for a device showing spin-split ABSs as reported in \cite{Tosi2019}. The thin lines correspond to the ballistic case and the thick lines to the presence of backscattering ($r\neq 0$). d) The four level system at a fixed phase value $\phi$ as a coupled two-spin system, the first (mode) qubit represented by Pauli matrices $\tau_i$ and the second (spin) qubit represented by Pauli matrices $\sigma_i$. Gate and flux drivings induce Rabi terms $\W_g\tau_x$ and $\W_f\tau_x$ respectively. Additionally, gate driving induces a $\alpha\W_g\tau_x\sigma_x$ interaction and a ferromagnetic term $g\tau_z\sigma_z$ exists between both qubits.}
\label{fig:levels2spin}
\end{figure}
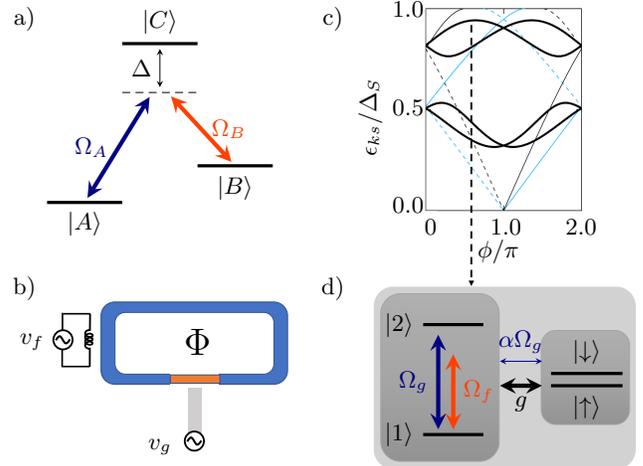

Recent experiments on weak links formed by semiconducting nanowires coupled to a superconducting 
resonator in a circuit QED geometry have enabled detection of their ABS spectrum \cite{Woerkom2017,Hays2018}, revealed their fine structure caused by spin-orbit interactions \cite{Tosi2019}, and offer the possibility to perform time resolved detection
of their population dynamics \cite{Hays2019}. While these results constitute a significant step forward,
these systems still have limited controllability and, more specifically, direct spin flip transitions can be hardly implemented in the absence of magnetic fields \cite{Park2017}. Thus, the coherent manipulation of the spin of a 
single quasiparticle trapped within these junctions requires special techniques which are still 
to be developed in the context of superconducting nanodevices. 

In the present work we propose to address these problems by adapting techniques which are common
in the field of quantum optics. A useful protocol for the control of population transfer between 
two states that lack a direct coupling is the {\em stimulated Raman adiabatic passage} (STIRAP) \cite{Gaubatz1990, Bergmann2019}. It consists 
in the application of two delayed overlapping pulses in Raman configuration with respect to a third state, as illustrated in Fig.\ref{fig:levels2spin}a). 
STIRAP is known to be robust against pulse shape modifications. This is in stark contrast with 
the simultaneous application of two fields in Raman configuration, which requires very accurate adjustment of the pulse area to a value of $\pi$, as well as careful adjustment of the Rabi frequency $\W$ of both fields to similar values. In addition, a large detuning $\Delta\gg\W$ is needed in order 
to suppress population transfer to the excited levels. This reduces the effective Rabi frequency, which translates into a transfer time that is proportional to $\frac{\Delta}{\W^2}$. This is in general a factor $\frac{\Delta}{\W}$ slower than STIRAP.

In the context of superconducting circuits, the STIRAP protocol was first proposed for Cooper-pair boxes \cite{Siewert2006} and experimentally demonstrated on transmons \cite{Kumar2016}. Unfortunately, STIRAP techniques cannot be applied directly to the Josephson weak link case. 
In the devices of Refs.\cite{Tosi2019,Hays2019} with a typical junction length of the order of 400 nm,
two set of spin-split ABSs appear within the superconducting gap. Their characteristic dispersion 
as a function of the superconducting phase difference $\phi$ is illustrated in Fig.\ref{fig:levels2spin}c). Thus, when 
$\phi$ is fixed to a value different from $0$ or $\pi$,  there are four states within the odd sector which can be connected by single particle transitions, instead of the three usually considered for STIRAP. 
These can be viewed as two coupled qubits, one with transition frequencies of the order of 10GHz, 
which is associated with different Andreev longitudinal modes and one with frequencies of the 
order of 1GHz which is associated with the spin degree of freedom. Ambient flux or charge noise 
fully mixes the spin degree of freedom \cite{Hays2019} and thus additional spin-polarization techniques are necessary to prepare an initially pure state which is required to apply the STIRAP protocol successfuly. 
We also notice that intra-doublet spin-flip transitions are in general suppressed compared to inter-doublet transitions \cite{Park2017} while spin-flip transitions between manifolds are allowed with an efficiency $\alpha(\phi)$ with respect to the spin-conserving ones. This efficiency has been experimentally observed to take values in a range between 0 and 1 both for gate driving \cite{Tosi2019,Metzger2020} and for flux driving \cite{Hays2019,Hays2020}. Although the parameter is kept in all subsequent derivations, for simplicity in the simulations we shall assume a combination of a purely spin-conserving drive ($\alpha=0$) as well as an $\alpha>0$ drive.

We first discuss a simplified modeling of the ABSs in these systems and then propose an implementation of a protocol analogous to STIRAP. We analyze the most appropriate preparation technique and produce an analytical model for its optimization. We show that the protocol is robust for a large range of driving strengths and spin decohering rates, thus providing an avenue for the control of ASQ regardless of device specifications.

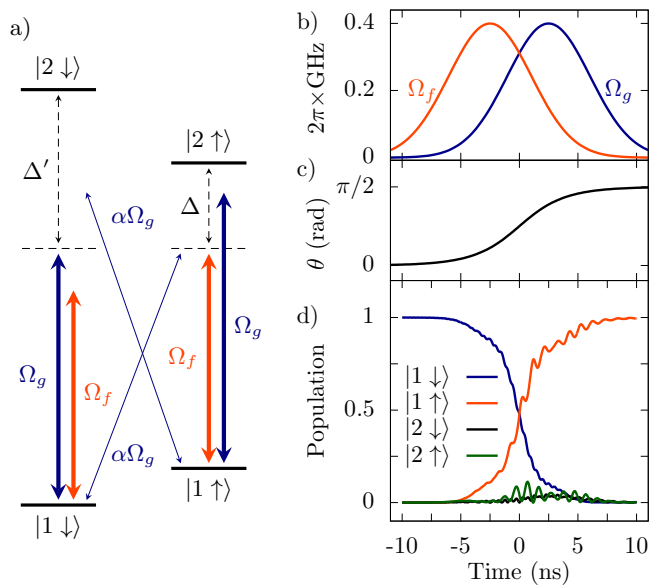
\begin{figure}[b]
\centerline{
    \begin{tikzpicture}[
      level/.style={very thick},
      virtual/.style={ultra thin,densely dashed},
      transthb/.style={ultra thick,blue!50!black,<->,shorten >=2pt,shorten <=2pt,>=stealth},
      transb/.style={ultra thin,blue!50!black,<->,shorten >=2pt,shorten <=2pt,>=stealth},
      transbl/.style={ultra thin,densely dashed,<->,shorten >=2pt,shorten <=2pt,>=stealth},
      transthr/.style={ultra thick,color=orange!50!red,<->,shorten >=2pt,shorten <=2pt,>=stealth},
    ]
    \draw[level] (0cm,-7.5em) -- (1cm,-7.5em) node[midway,below] {$\ket{1\da}$};
    \draw[level] (2cm,-6em) -- (3cm,-6em) node[midway,below] {$\ket{1\ua}$};
    \draw[level] (0cm,9.5em) -- (1cm,9.5em) node[midway,above] {$\ket{2\da}$};
    \draw[level] (2cm,6.5em) -- (3cm,6.5em) node[midway,above] {$\ket{2\ua}$};
    \draw[virtual] (2cm,3em) -- (3cm,3em) node[midway,above] {};
    \draw[virtual] (1cm,3em) -- (0cm,3em) node[midway,above] {};
    \draw[transb] (.85cm,-7.5em) -- (2.15cm,3em) node[]  at (1.5cm,4.25em) {$\alpha\W_g$};
    \draw[transthb] (.5cm,-7.5em) -- (.5cm,3em) node[midway,left] {$\W_g$};
    \draw[transb] (2.15cm,-6em) -- (.85cm,5.5em) node[] at (1.5cm,-5.5em) {$\alpha\W_g$};
    \draw[transthb] (2.7cm,-6em) -- (2.7cm,5.5em) node[midway,right] {$\W_g$};
    \draw[transthr] (2.5cm,-6em) -- (2.5cm,3em) node[midway,left] {$\W_f$};
    \draw[transthr] (.7cm,-7.5em) -- (.7cm,1.5em) node[midway,right] {$\W_f$};
    \draw[transbl] (2.5cm,3em) -- (2.5cm,6.5em) node[midway,left] {$\Delta$};
    \draw[transbl] (0.5cm,3em) -- (0.5cm,9.5em) node[midway,left] {$\Delta'$};
    \node[draw=none,rectangle]  at (0cm,12em) {a)};
    \node[] (russell) at (6cm,1em)
    {\includegraphics[scale=0.9]{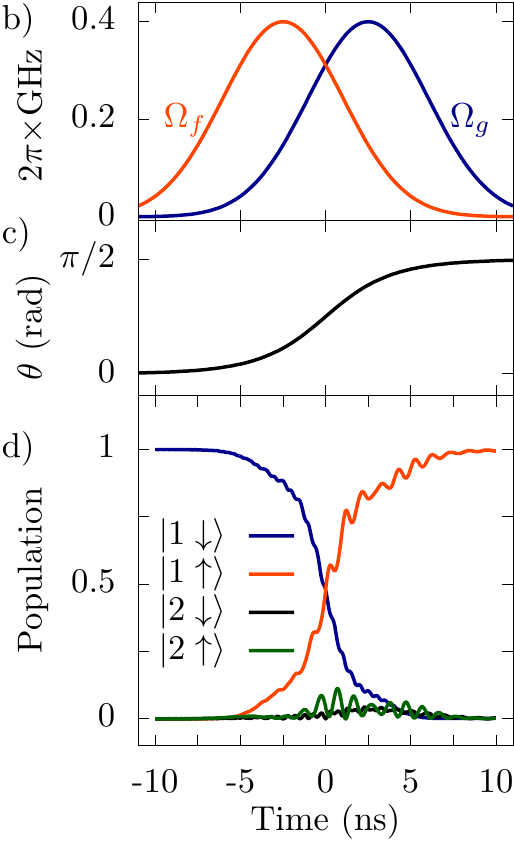}};
    \end{tikzpicture}
}
\caption{a) Schematic level diagram with available couplings generated by gate AC driving (blue) and magnetic flux AC driving (orange). The frequencies $\w_g$ and $\w_f$ are chosen such that the Raman resonance condition Eq.(\ref{eq:Raman}) is fulfilled. b) Rabi frequencies $\W_g$ and $\W_f$ as a function of time for $\alpha=1$. Otherwise, the intensity of the gate pulse has to be enhanced by $\alpha^{-1}$ to achieve transfer. c) Value of $\theta=\arctan(\alpha\W_g/\W_f)$ as a function of time for the pulses in b). d) Dynamics of the occupation probabilities of states $\ket{1\da}$, $\ket{1\ua}$, $\ket{2\da}$ and $\ket{2\ua}$ under the effect of the pulses in b) and for $\alpha=1$, $\Delta=0$ and $\Delta'=2\pi \times 2{\rm GHz}$.}
\label{fig:STIRAP4}
\end{figure}

{\em System Modeling ---} 
In order to obtain a simple model for the ABSs in the nanowire Josephson junction we consider the ballistic case with a single occupied subband, characterized by two modes $\ket{m_{1,2}}$ with different Fermi velocities $v_{1,2}$ \cite{Tosi2019,SM}. The low energy effective Hamiltonian can be written in this basis as
\bea
\label{eq:Hphys}
H_{eff}&=&\sum_{j=1,2}\frac{\Delta_S}{1+\lambda_j}\frac{\pi-(-1)^j\phi \sigma_z}{2}\prj{m_j}\\
& + & r\Delta_S \left(\ket{m_1}\bra{m_2} + H.c.\right),  \nonumber
\eea
where $\Delta_S$ is the superconducting gap, $\lambda_j=\Delta_S L/\hbar v_j$ where $L$ is the junction length, $r$ the reflection coefficient of the weak link and $\sigma_z$ the Pauli matrix for the pseudospin degree of freedom in the basis spanned by states $\ket{\ua}$ and $\ket{\da}$. More details are provided in the supplementary information \cite{SM}.

After diagonalization, it can be written as \cite{SM}
\bea
H_{eff}  &= &\frac{\w}{2} \tau_z - \frac{g}{2}  \tau_z\sigma_z + \frac{\nu}{2} \sigma_z + E_0 \\
&= &\sum_{\substack{k=1,2 \\ s=\ua,\da}}\e_{ks}\prj{ks} \nonumber
\label{eq:H}
\eea
where we now define $\tau_z$ as the Pauli matrix for the Andreev longitudinal mode degrees of freedom in the basis spanned by states $\ket{1}$ and $\ket{2}$ with an attractive interaction term of strength $g$ that inverts the spin orientation in the upper level and a global energy offset $E_0$. Note that the degrees of freedom have become rotated with respect to Eq.(\ref{eq:Hphys}).

We consider a gate AC-driving at frequency $\w_g$ that produces pseudospin-conserving transitions of strength $\W_g$ and pseudospin-flipping transitions of weaker strength $\alpha \W_g$
\be
H_g(t) = \W_g\cos(\w_g t) \left(1 + \alpha \sigma_x  \right)\tau_x,
\ee
and an additional flux AC-driving at frequency $\w_f$ that produces pseudospin-conserving transitions of strength $\W_f$ and no pseudospin-flipping transitions ($\alpha=0$)
\be
H_f(t) = \W_f\cos(\w_f t)\tau_x.
\ee
A layout of the available degrees of freedom and couplings is shown in Fig.\ref{fig:STIRAP4}a).

{\em STIRAP Implementation ---} As shown in Fig.\ref{fig:STIRAP4}a), we consider a four level system unlike the typical STIRAP setting of three levels (see Fig.\ref{fig:levels2spin}a)). In order to implement a STIRAP protocol between the initial state $\ket{1\da}$ and the final state $\ket{1\ua}$, the following Raman resonance condition is imposed
\be
\Delta \equiv
\left(\e_{2\ua} - \e_{1\da} \right) - \w_g =
\left(\e_{2\ua} - \e_{1\ua} \right) - \w_f
\label{eq:Raman}
\ee
such that a {\em rotating wave approximation} (RWA) in the relevant interaction picture reveals the transitions
\bea
H'_g +H'_f &\simeq&  \frac{\W_f}{2} \ket{1\ua}\bra{2\ua}
  + \frac{\alpha\W_g}{2} \ket{1\da}\bra{2\ua} \nonumber \\
&&+\frac{\W_g}{2} \ket{1 \da}\bra{2 \da}
  + \mathrm{H.c.}
\eea
including the wanted $\Lambda$-coupling that satisfies the Raman condition and an unwanted vertical coupling detuned by $\Delta'=\left(\e_{2\da} - \e_{1\da} \right) - \w_g$. It is important to notice that, beyond these three couplings, the RWA neglects three additional unwanted non-resonant couplings as shown in Fig.\ref{fig:STIRAP4}a)). In particular, the coupling between $\ket{1\ua}$ and $\ket{2\ua}$ of strength $\W_g$ may be even more detrimental due to its proximity to resonance. These couplings can only be captured in a full time-dependent numerical simulation of the sort performed in this work. Their effect is in general suppressed with the choice $\Delta=0$, which is the default choice in the remainder of the discussion unless otherwise stated.

\begin{figure}[b]
\centerline{
    \begin{tikzpicture}[
      level/.style={very thick},
      virtual/.style={ultra thin,densely dashed},
      transthb/.style={ultra thick,blue!50!black,<->,shorten >=2pt,shorten <=2pt,>=stealth},
      transb/.style={ultra thin,blue!50!black,<->,shorten >=2pt,shorten <=2pt,>=stealth},
      transbl/.style={ultra thin,densely dashed,<->,shorten >=2pt,shorten <=2pt,>=stealth},
      transthr/.style={ultra thick,color=orange!50!red,<->,shorten >=2pt,shorten <=2pt,>=stealth},
    ]
    \draw[level] (1cm,-3em) -- (0cm,-3em);
    \draw[level] (2cm,-1.5em) -- (3cm,-1.5em);
    \draw[level] (1cm,5.5em) -- (0cm,5.5em);
    \draw[level] (2cm,2.5em) -- (3cm,2.5em);
    \draw[virtual] (2cm,4em) -- (3cm,4em) node[midway,above] {};
    \draw[transb] (.85cm,-3em) -- (2.15cm,4em) node[]  at (1.45cm,4.2em) {$\alpha\W_g$};
    \draw[transthb] (.5cm,-3em) -- (.5cm,4em) node[midway,left] {$\W_g$};
    \draw[transb] (2.15cm,-1.5em) -- (.85cm,5.5em) node[] at (1.4cm,-2em) {$\alpha\W_g$};
    \draw[transthb] (2.5cm,-1.5em) -- (2.5cm,5.5em) node[] at (2.8cm,.5em) {$\W_g$};
    \draw[transbl] (2.9cm,4em) -- (2.9cm,2.5em) node[midway,right] {$-\Delta$};
    \node[draw=none,rectangle]  at (-.3cm,6em) {a)};
          \node[] (wazne) at (3.7cm,-12em)
    {\includegraphics[scale=0.9]{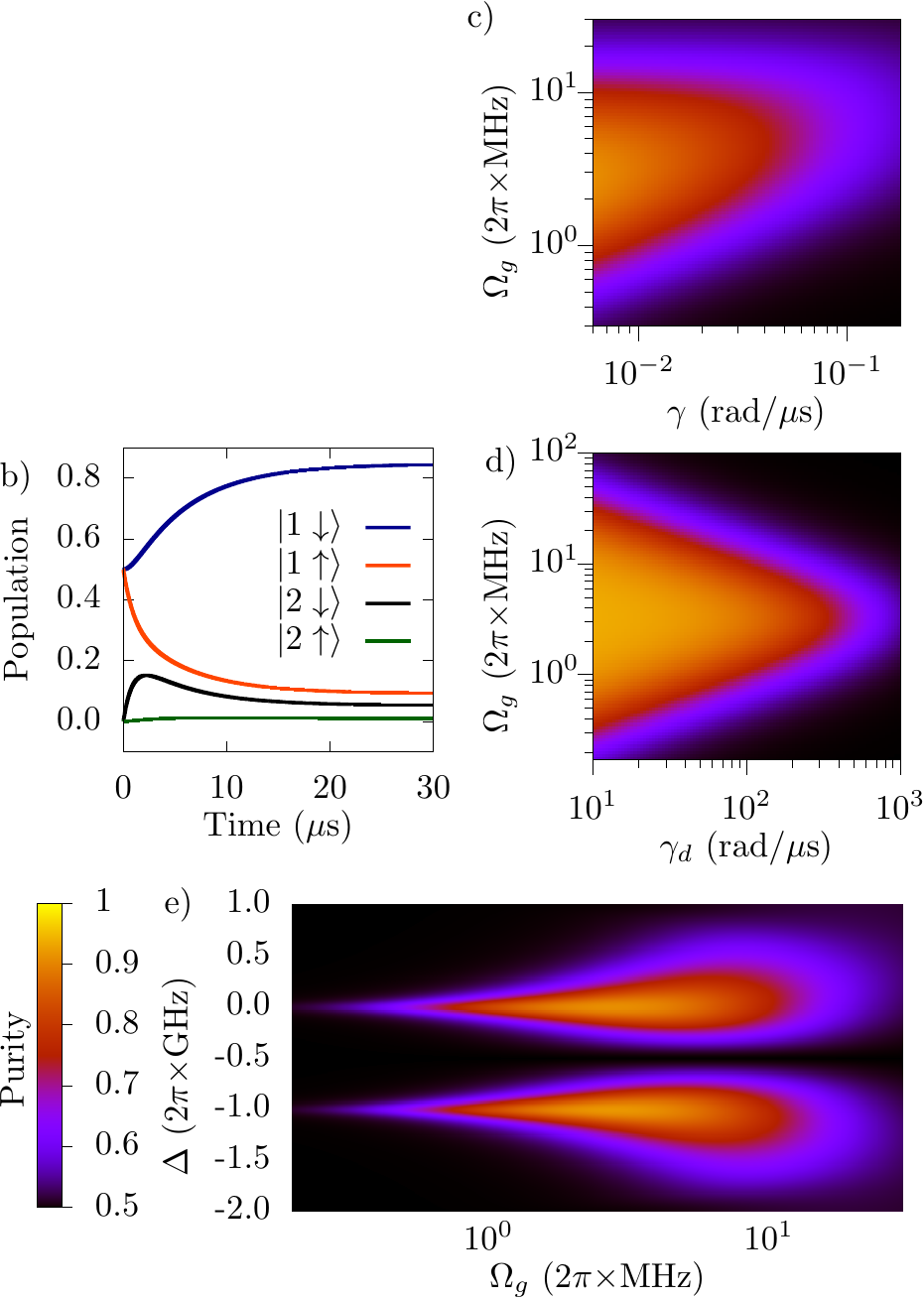}};
    \end{tikzpicture}
}
\caption{a) Resonance condition $\w_g=\e_{2\da}-\e_{1\ua}$ for the preparation of state $\ket{1\da}$ with a single gate pulse (Note the difference with the arrangement of Fig.\ref{fig:STIRAP4}a). b) Population dynamics under the effect of a continuous gate pulse of strength $\W_g=2\pi\times 3\,\mathrm{MHz}$  for $\gamma =20\,\mathrm{rad/ms}$ and $\gamma_d =100\,\mathrm{rad/\mu s}$. c) Purity of the steady state of the spin degrees of freedom as a function of the driving Rabi frequency $\W_g$ and the spin mixing rate $\gamma$ for $\gamma_d=100\,\mathrm{rad/\mu s}$.  d) Purity of the steady state of the spin degrees of freedom as a function of the driving Rabi frequency $\W_g$ and the decoherence rate $\gamma_d$ for $\gamma= 10\,\mathrm{rad/ms}$. e) Purity of the steady state of the spin degrees of freedom as a function of the driving Rabi frequency $\W_g$ and the detuning $\Delta$ for $\gamma =5\,\mathrm{rad/ms}$ and $\gamma_d =100\,\mathrm{rad/\mu s}$. Purity values may range between $1$ (pure state) and $0.5$ (fully mixed 2-level system).}
\label{fig:cooling}
\end{figure}

The STIRAP protocol consists of the introduction of a slow modulation of the parameters $\W_f$ and $\W_g$
\be
\W_f (t) = \alpha A e^{-\left( \frac{t-t_f}{T} \right)^2},\quad \W_g (t) = A e^{-\left( \frac{t-t_g}{T} \right)^2}, 
\label{eq:pulses}
\ee
at times $t_f<t_g$ with pulse amplitudes $\alpha A$, $A$ and widths $T$.
Under the time-dependent transformation of the lower states into a bright $\ket{B}$ and a dark $\ket{D}$ states defined by
\bea
\ket{B}&=&\frac{1}{\W}\ket{1}\left(\alpha\W_g\ket{\da}+\W_f\ket{\ua}\right),\nonumber \\
\ket{D}&=&\frac{1}{\W}\ket{1}\left(\W_f\ket{\da}-\alpha\W_g\ket{\ua}\right),
\eea
with $\W=\sqrt{\W_f^2 + \alpha^2\W_g^2}$ the root mean square Rabi frequency, the effective Hamiltonian simplifies to the form
\bea
H(t)&=&H'_{eff}+H'_g +H'_f\\
\nonumber &\simeq& \left(\Delta\prj{\ua} + \Delta'\prj{\da}\right)\prj{2}\\
&& + \frac{\W}{2} \left( \ket{2\ua}\bra{B} + H.c.\right) + i \dot{U}U^\dagger,\nonumber
\eea
where the time-dependent transformation $U= \ket{D}\bra{1\da} + \ket{B}\bra{1\ua} + \sum_{\sigma=\da,\ua}\prj{2\sigma}$ introduces the term
\be
i\dot{U}U^\dagger=i\dot\theta\ket{B}\bra{D}+H.c.
\label{eq:Hna}
\ee
with $\tan\theta=\alpha\frac{\W_g}{\W_f}$ so that $\dot\theta=\alpha\frac{\dot\W_g\W_f-\dot\W_f\W_g}{\W^2}$. For very slow (adiabatic) variations of the Rabi frequency, the dark state $\ket{D}$ remains decoupled. The mechanism of STIRAP is straightforward from here. A system is initially prepared in $\ket{1\da}$, which becomes the dark state under the effect of the first pulse, and can be adiabatically transformed into any superposition with $\ket{1\ua}$ by an appropiate choice of the timing of $\theta$. The condition of adiabaticity can be expressed as
\be
 \frac{\dot\theta}{\W} \ll 1  \label{eq:AdCond}
\ee
at all times $t$.

We present an implementation of the protocol in Fig.\ref{fig:STIRAP4}. As an example, following \cite{Hays2019} we may assume a spectrum with non-degenerate transitions such that 
\bea
\e_{1\ua}-\e_{1\da}&=& 2\pi\times 1 \,\mathrm{GHz}, \nonumber \\
\e_{2\da}-\e_{1\da}&=& 2\pi\times 13 \,\mathrm{GHz}, \nonumber \\
\e_{2\ua}-\e_{1\da}&=& 2\pi\times 11 \,\mathrm{GHz}, \label{eq:spec}
\eea
the choice $\Delta=0$ imposes $\w_g=2\pi\times 11 \,\mathrm{GHz}$, $\w_f=2\pi\times 10\,\mathrm{GHz}$,  and $\Delta'=2\pi\times 2\,\mathrm{GHz}$. Assuming no suppression of spin-flipping transitions $\alpha=1$, a choice of timings $t_g=2T=-t_f=2.5\,\mathrm{n s}$ imposes a minimum pulse strength $A=2\pi\times 400\,\mathrm{MHz}$ in order for Eq.(\ref{eq:AdCond}) to be satisfied. The population transfer corresponding to such pulses (Fig.\ref{fig:STIRAP4}b)-c)) is shown in Fig.\ref{fig:STIRAP4}d). Note a rather small transfer of population to level $\ket{2\ua}$ due to the presence of the strong non-resonant coupling between $\ket{1\ua}$ and $\ket{2\ua}$ generated by the gate driving. Our parametric choice prevents this effect from affecting the success of STIRAP.

{\em Dissipation and state preparation ---} As investigated in \cite{Hays2019, Hays2020}, the states have both a limited lifetime and coherence time. In their sample, two pseudospin-conserving decay channels between excited and ground levels are found of rates $\Gamma$ ranging between $250\,\mathrm{rad/ms}$ and $333\,\mathrm{rad/ms}$. Additionally, a pseudospin mixing term of rate $\gamma$ drives the system into a maximal mixture of states $\ket{\da}$ and $\ket{\ua}$. The numeric value has been investigated in \cite{Hays2019} with an inconclusive range between $\simeq24.4\,\mathrm{rad/ms}$ and $\simeq0.4\,\mathrm{rad/ms}$, as discussed in \cite{SM}. Additionally, the coherence of the spin degrees of freedom decays at a rate $\gamma_d$ in the order of $100\,\mathrm{rad/\mu s}$ \cite{Hays2020}. We model these decoherence effects by means of the master equation
\bea
\frac{d}{dt}\rho &=&\left[H(t),\rho\right]+\frac{\gamma_d}{2}\mathcal{D}\left[\tau_z\right](\rho) + \frac{\gamma_d}{2}\mathcal{D}\left[\sigma_z\right](\rho) \label{eq:ME2}\\
&+&\frac{\Gamma}{2}\mathcal{D}\left[\tau_-\right](\rho) + \frac{\gamma}{2}\mathcal{D}\left[\sigma_-\right](\rho) + \frac{\gamma}{2}\mathcal{D}\left[\sigma_+\right](\rho) , \nonumber
\eea
where $\tau_-=\ket{1}\bra{2}$, $\sigma_-=\ket{\da}\bra{\ua}$, $\sigma_+=\left(\sigma_-\right)\x$ and with the Lindbladian superoperator $\mathcal{D}\left[ s\right](\rho)=2 s\rho s\x - s\x s \rho - \rho s\x s$. Note that we assume a similar coherence lifetime of superpositions of $\ket{1}$ and $\ket{2}$ states in line with the findings in \cite{Hays2020}.

Due to these processes, the system is initially in the fully mixed state $\rho(0)=1/2\prj{1}\sum_{s=\da,\ua}\prj{s}$, which imposes the need to prepare the sample in a state of low entropy for any control protocol. Fortunately, the gate pulse in the STIRAP arrangement (Fig.\ref{fig:STIRAP4}a)) automatically produces an optical pumping effect that depletes the state $\ket{1\da}$, similar to the case shown in \cite{Hays2020}. For an appropriate choice of $\Delta$ (see Fig.\ref{fig:cooling}a)), the state $\ket{1\ua}$ may be depleted instead (Fig.\ref{fig:cooling}b)). This is due to the resonance with the $\ket{1\ua}-\ket{2\da}$ transition, which pumps population from $\ket{\ua}$ to $\ket{\da}$ that then preferentially decays to $\ket{1\da}$.

\begin{figure}[b]
\includegraphics[width=\columnwidth]{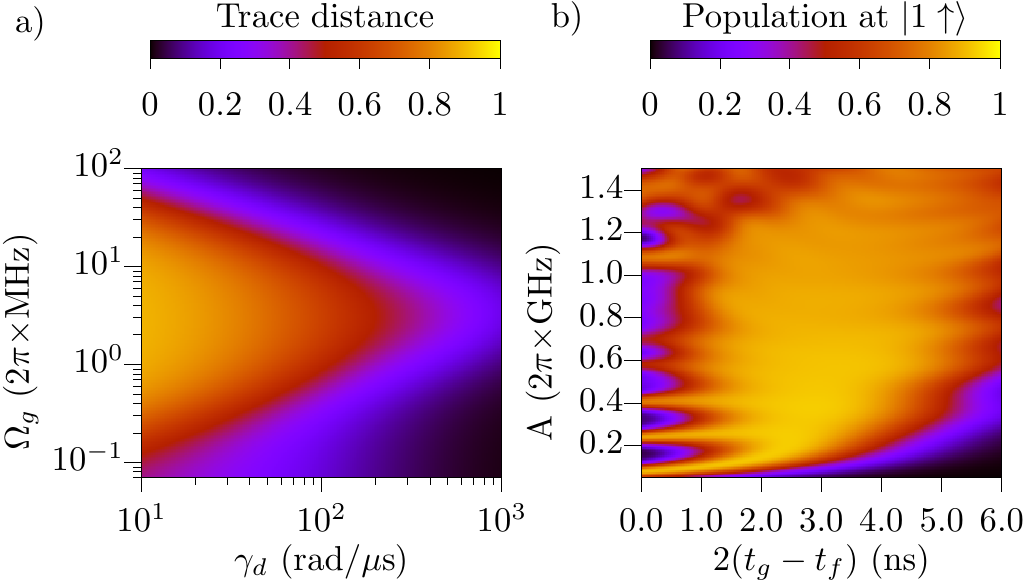}
\caption{a) Trace distance between the spin state before and after the STIRAP protocol as a function of the Rabi frequency $\W_g$ of the preparation pulse and the decoherence rate $\gamma_d$ for $\gamma=  10\,\mathrm{rad/ms}$. STIRAP pulses as in Fig.\ref{fig:STIRAP4}b). b) Population transferred from state $\ket{1\da}$ to state $\ket{1\ua}$ after application of the STIRAP protocol as a function of the Rabi frequency $A$ and the pulse separation $2(t_g-t_f)$ for $T=5\,\mathrm{ns}$, a spin mixing rate $\gamma=10\,\mathrm{rad/ms}$ and a decoherence rate $\gamma_d= 100\,\mathrm{rad/\mu s}$.}
\label{fig:TrDist}
\end{figure}

The effect of the polarizing pulse may be described in terms of the rate equation \cite{SM}
\be
\frac{d}{dt}
\begin{pmatrix}
p_\da\\
p_\ua
\end{pmatrix}
=
\begin{pmatrix}
-\gamma_\ua-\gamma & \gamma_\da+\gamma \\
\gamma_\ua+\gamma & -\gamma_\da-\gamma
\end{pmatrix}
\begin{pmatrix}
p_\da\\
p_\ua
\end{pmatrix},
\ee
For an optimal choice of parameters, $\gamma_\ua \simeq 0 $ and $\gamma_\da \simeq\Gamma$, so that the population inversion $\Gamma/(2\gamma+\Gamma)$ is reached. This is only going to approach 1 for $\gamma\ll\Gamma$ (see Fig.\ref{fig:cooling}b), and this becomes a crucial criterion to select the parametric region for the implementation of STIRAP. The choice of $\W_g$ is additionally affected by the decoherence rate $\gamma_d$, as shown in Fig.\ref{fig:cooling}c). Finally, it is equally possible to address the detuning value $\Delta=0$ in order to prepare the state $\ket{1\ua}$ with similar efficiency.

{\em Global efficiency of the protocol ---} The final success of state transfer from $\ket{1\da}$ into $\ket{1\ua}$ after the polarization pulse into $\ket{1\da}$ and the STIRAP protocol can be measured in terms of the trace distance $Tr[\sqrt{(\rho_0-\rho_t)^2}]$ between the spin density matrices before $\rho_0$ and after $\rho_t$ the STIRAP protocol. Its maximum is 1 for orthogonal states, while it vanishes for equal states. For a 2-level system, it represents an appropriate measure of the success of the transfer for mixed states. This is shown in Fig.\ref{fig:TrDist}a). The transfer is successful for a large range of preparation pulse strengths and decoherence rates, but it is mostly dependent on the purity of the initial state, as one can deduce from comparing Figs. \ref{fig:TrDist}b) and \ref{fig:cooling}d). At the same time, variations in the overall Rabi frequency $A$ and the separation betweeen pulses $t_g-t_f$ do not substantially affect the final transfer efficiency, as shown in Fig.\ref{fig:TrDist}b). On the contrary, application of continuous Raman pulses (corresponding to the left edge of the same figure) is highly sensitive to pulse strength, whereas STIRAP is robust over a large region. This makes this proposal suitable for varying design parameters.

{\em Conclusions ---} We have proposed a means to the manipulation of the pseudospin degrees of freedom of Andreev bound states that include a polarization pulse and subsequent coherent manipulation with the STIRAP technique. We have shown that the protocol could be efficient even for the stringent conditions of actual present experiments in which coherent times are rather short and is in line with current efforts to develop an ASQ by several experimental groups.

{\em Acknowledgments} ---
A.L.Y., J.C.  acknowledge European Union’s Horizon 2020 research and innovation programme for financial support grant no. 828948 (AndQC) and QuantERA project no. 127900 (SuperTOP).
J.C. acknowldeges the support from {\em Ministerio de Ciencia, Innovaci\'on y Universidades} (Spain) (``Beatriz Galindo'' Fellowship BEAGAL18/00081).
A.L.Y. acknowledges support by Spanish MICINN through grants FIS2017-84860-R and through the “María de Maeztu” Programme for Units of Excellence in R\&D (Grant No. MDM-2014-0377).
M.H. and V.F. acknowldege support by the US Office of Naval Research (N00014-16-1-2270) and by the US Army Research Office (W911NF-18-1-0020, W911NF-18-1-0212 and W911NF-16-1-0349).

\newpage
\newpage

\onecolumngrid

\section{Supplemental Information}

\section{Effective Model derivation}
{\it Ballistic case:} As shown in Ref. \cite{sTosi2019} a $\sim$ 400 nm long nanowire Josephson junction typically exhibits 
two sets of spin-split ABSs for a single channel situation. We take the ballistic case
analyzed in Refs. \cite{sPark2017,sTosi2019} as a reference to describe the type of phase dependence and the spin splitting exhibited by these states. For energies small compared to the superconducting gap, $\Delta_S$, the lowest energy states can be approximated as \cite{sTosi2019}
\begin{eqnarray*}
\epsilon_{m_1,\uparrow}(\phi) &=& \frac{\Delta_S}{1+\lambda_1}\frac{\pi+\phi}{2} \nonumber\\
\epsilon_{m_1,\downarrow}(\phi) &=& \frac{\Delta_S}{1+\lambda_1}\frac{\pi-\phi}{2} \nonumber\\
\epsilon_{m_2,\uparrow}(\phi) &=& \frac{\Delta_S}{1+\lambda_2}\frac{\pi-\phi}{2} \nonumber\\
\epsilon_{m_2,\downarrow}(\phi) &=& \frac{\Delta_S}{1+\lambda_2}\frac{\pi+\phi}{2} \;, \nonumber\\
\end{eqnarray*}
where $m_{1,2}$ indicate the modes with different Fermi velocities and $\uparrow,\downarrow$ the corresponding pseudospin index. On the other hand $\lambda_{j=1,2}=L\Delta_S/(\hbar v_j)$, where $L$ is the junction length and
\begin{equation}
v_j = \frac{\hbar k_{Fj}}{m^*} + (-1)^j \frac{\alpha_{SO}\left(E^{\perp}_1/2-(-1)^j \alpha_{SO} k_{Fj}\right)}{\hbar\sqrt{\left(E^{\perp}_1/2 - (-1)^j\alpha_{SO} k_{Fj}\right)^2 + \eta^2}},
\label{Velocity}
\end{equation}
where $E_1^{\perp}$ is the ground state energy of the transverse modes, $\alpha_{SO}$ the strength of the spin-orbit coupling, $\eta$ the spin-flipping coupling strength to the first excited transverse mode and $k_{Fj}$ are the Fermi wave vectors satisfying $E_s(k_{Fj})=\mu$,
with
\begin{equation}
E_s(k_x) = \frac{\hbar^2 k^2_x}{2 m^*} + \frac{E^{\perp}_1 +E^{\perp}_2}{2} 
- \sqrt{\left(  \frac{E^{\perp}_1 - E^{\perp}_2}{2} -s \alpha_{SO} k_x\right)^2 + \eta^2}, 
\label{Dispersion}
\end{equation} 
where $E_2^{\perp}$ is the energy of the first excited transverse mode and $s=\mp 1$ corresponds to the fast and the slow modes respectively. Introducing the Pauli matrix $\sigma_z=\prj{\ua}-\prj{\da}$ and states $\ket{m_1}$ and $\ket{m_2}$, the effective Hamiltonian in the ballistic case can then be written as
\begin{equation}
H_0 = \frac{\Delta_S}{1+\lambda_1}\frac{\pi+\phi \sigma_z}{2}|m_1\rangle \langle m_1|  + \frac{\Delta_S}{1+\lambda_2}\frac{\pi-\phi \sigma_z}{2} |m_2\rangle \langle m_2|   \;.
\end{equation}

{\it Beyond ballistic case:} The main effect of elastic scattering is to introduce coupling between ballistic states and thus can be modelled as an extra term in the Hamiltonian coupling the modes without mixing the pseudospin, which yield the model Hamiltonian given in Eq. (1) of the main text 
\begin{equation}
H_{eff}= H_0 + \Delta_S r\left(\ket{m_1}\bra{m_2}+H.c.\right).
\end{equation}
Notice that this model is valid for $\phi \ll \pi$ (around $\pi$ the coupling with other modes should be taken into account). This Hamiltonian may be analytically diagonalized since
\be
\ave{\ua\left|H_{eff}\right|\ua}=\Delta_S
\begin{pmatrix}
 \frac{1}{1+\lambda_1}\frac{\pi +\phi}{2} & r \\
r &  \frac{1}{1+\lambda_2}\frac{\pi-\phi}{2}
\end{pmatrix}
\ee
and
\be
\ave{\da\left|H_{eff}\right|\da}=\Delta_S
\begin{pmatrix}
 \frac{1}{1+\lambda_1}\frac{\pi -\phi}{2} & r \\
r &  \frac{1}{1+\lambda_2}\frac{\pi +\phi}{2}
\end{pmatrix}.
\ee
This produces eigenstates
\begin{eqnarray*}
\ket{1\uparrow} &=& \left(\cos\theta_\ua\ket{m_1}+\sin\theta_\ua\ket{m_2}\right)\ket{\ua} \nonumber\\
\ket{2\uparrow} &=& \left(-\sin\theta_\ua\ket{m_1}+\cos\theta_\ua\ket{m_2}\right)\ket{\ua} \nonumber\\
\ket{1\downarrow} &=&  \left(\cos\theta_\da\ket{m_1}+\sin\theta_\da\ket{m_2}\right)\ket{\da} \nonumber\\
\ket{2\downarrow} &=& \left(-\sin\theta_\da\ket{m_1}+\cos\theta_\da\ket{m_2}\right)\ket{\da} \;, \nonumber\\
\end{eqnarray*}
with 
\bea
\tan2\theta_\ua&=&4r \frac{\left(1+\lambda_1\right)\left(1+\lambda_2\right)}{\lambda_2\left(\pi+\phi\right)-\lambda_1\left(\pi-\phi\right)+2\phi }\nonumber\\
\tan2\theta_\da&=&4r \frac{\left(1+\lambda_1\right)\left(1+\lambda_2\right)}{\lambda_2\left(\pi-\phi\right)-\lambda_1\left(\pi+\phi\right)-2\phi}\nonumber
\eea
and eigenvalues
\begin{eqnarray*}
\epsilon_{1 \uparrow}(\phi) &=& \frac{\Delta_S}{4}\left[\frac{\pi +\phi}{1+\lambda_1}+\frac{\pi -\phi}{1+\lambda_2}-\sqrt{\left(\frac{\pi +\phi}{1+\lambda_1}-\frac{\pi -\phi}{1+\lambda_2}\right)^2+16 r^2}\right] \nonumber\\
\epsilon_{2 \uparrow}(\phi) &=&  \frac{\Delta_S}{4}\left[\frac{\pi +\phi}{1+\lambda_1}+\frac{\pi -\phi}{1+\lambda_2}+\sqrt{\left(\frac{\pi +\phi}{1+\lambda_1}-\frac{\pi -\phi}{1+\lambda_2}\right)^2+16 r^2}\right] \nonumber\\
\epsilon_{1 \downarrow}(\phi) &=& \frac{\Delta_S}{4}\left[\frac{\pi -\phi}{1+\lambda_1}+\frac{\pi +\phi}{1+\lambda_2}-\sqrt{\left(\frac{\pi -\phi}{1+\lambda_1}-\frac{\pi +\phi}{1+\lambda_2}\right)^2+16 r^2}\right] \nonumber\\
\epsilon_{2 \downarrow}(\phi) &=& \frac{\Delta_S}{4}\left[\frac{\pi -\phi}{1+\lambda_1}+\frac{\pi +\phi}{1+\lambda_2}+\sqrt{\left(\frac{\pi -\phi}{1+\lambda_1}-\frac{\pi +\phi}{1+\lambda_2}\right)^2+16 r^2}\right] \;, \nonumber\\
\end{eqnarray*}
so that
\be
H_{eff}=\sum_{\substack{k=1,2 \\ s=\ua,\da}}\e_{ks}\prj{ks}.
\ee
This four level system can be understood as a pair of two level systems with bases $\left\{\ket{1},\ket{2}\right\}$ and $\left\{\ket{\ua},\ket{\da}\right\}$ and Pauli matrices $\tau_z=\prj{2}-\prj{1}$ and $\sigma_z=\prj{\ua}-\prj{\da}$ such that
\be
H_{eff}=\frac{\w}{2} \tau_z - \frac{g}{2}  \tau_z\sigma_z + \frac{\nu}{2} \sigma_z + E_0,
\ee
with
\begin{eqnarray*}
E_0 &=& \frac{1}{4}\sum_{\substack{k=1,2 \\ s=\ua,\da}}\e_{ks} = \frac{\pi\Delta_S}{4}\left(\frac{1}{1+\lambda_1}+\frac{1}{1+\lambda_2}\right)\;, \nonumber\\
\nu = \epsilon_{2,\ua} + \epsilon_{1,\ua}  - 2 E_0 &=& \frac{\phi \Delta_S}{2}\frac{\lambda_2-\lambda_1}{\left(1+\lambda_1\right)\left(1+\lambda_2\right)}, \nonumber\\
\w = \epsilon_{2\ua}+\epsilon_{2,\da} - 2 E_0 &=&  \frac{\Delta_S}{4}\sqrt{\left(\frac{\pi - \phi}{1+\lambda_1}-\frac{\pi +\phi}{1+\lambda_2}\right)^2+16 r^2} + \frac{\Delta_S}{4}\sqrt{\left(\frac{\pi +\phi}{1+\lambda_1}-\frac{\pi -\phi}{1+\lambda_2}\right)^2+16 r^2}, \nonumber\\
g = \epsilon_{2,\da} + \epsilon_{1,\ua} -2E_0 &=&   \frac{\Delta_S}{4}\sqrt{\left(\frac{\pi - \phi}{1+\lambda_1}-\frac{\pi +\phi}{1+\lambda_2}\right)^2+16 r^2} -  \frac{\Delta_S}{4}\sqrt{\left(\frac{\pi +\phi}{1+\lambda_1}-\frac{\pi -\phi}{1+\lambda_2}\right)^2+16 r^2}.  \nonumber\\
\end{eqnarray*}

\begin{figure}
\begin{center}
 \includegraphics[width=.7\columnwidth]{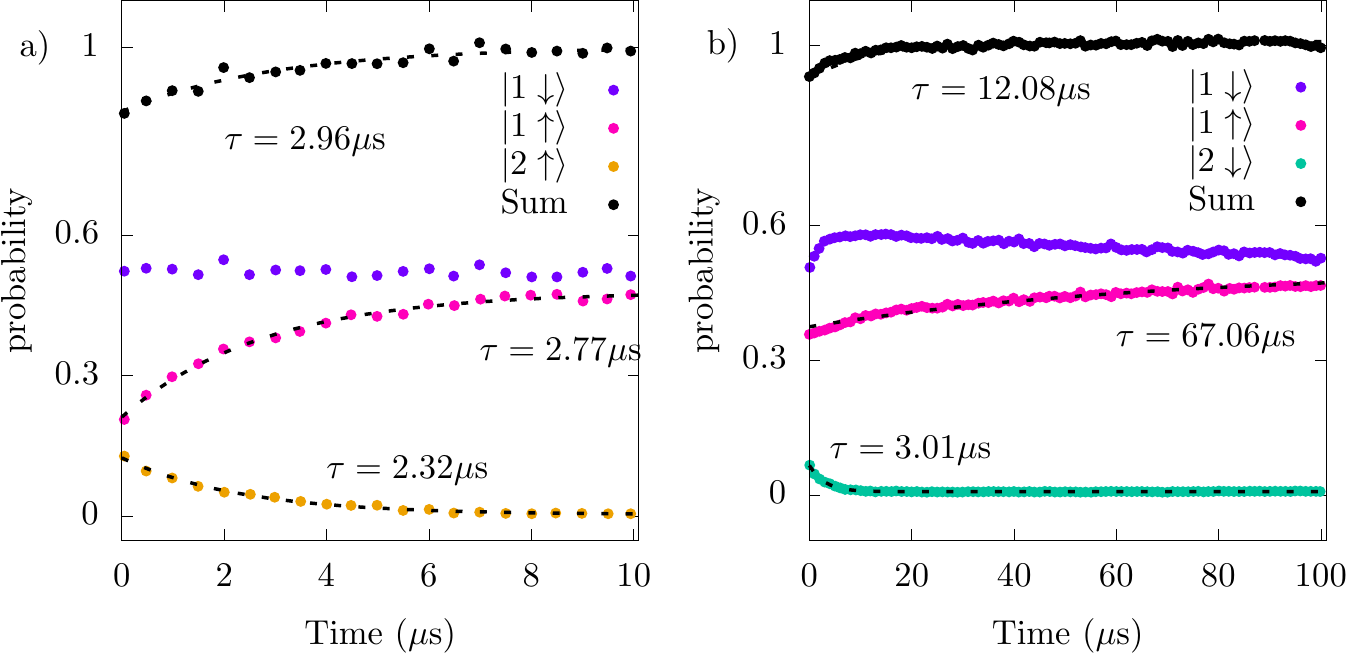}
   \end{center}
\caption{Probability of detected states as a function of time for two initial preparations (a) and (b) as in Fig.3 (d) and (g) of \cite{Hays2019}. Each data series is fit to an independent exponential decay, each providing disparate $\tau$ values. The probability of detecting a state (black dots) increases with time as the dynamics of the system (other colors) diminish.}
\label{fig:Fits1}
\end{figure}

\section{Spin-Mixing Rate}

The values of the parameters in the main text are taken from real experimental devices such as \cite{sTosi2019} and \cite{sHays2019}. Nevertheless, the value for the spin-mixing rate $\gamma$ when there exists simultaneous driving is ambiguous and leaves some space for discussion.

The method of datapoint assignation described in \cite{Hays2019} leaves out a set such that we can distinguish between detected and undetected states. We enlarge the probability space accordingly: beyond the four relevant physical states $s\in\{1\da,1\ua,2\da,2\ua\}$, the probability distribution $p(s,d)$ is defined over an additional detected/undetected space $d\in\{1,0\}$. The data in Fig.3 of \cite{Hays2019} corresponds to the distribution $p(s,1)$, which appears unnormalized since there are a number of undetected cases and $\sum_i p(i,1)=1-\sum_i p(i,0)<1$. We define $p_d(1)=\sum_i p(i,1)$ and plot it along the data of Fig.3(d) and Fig.3(g) of \cite{sHays2019} in the fits of Fig.\ref{fig:Fits1} (a) and (b), which evidence dynamics in the $d$ subspace.

We want to extract the rate equation that best reproduces the evolution of the probability distribution $p_s(s)=\sum_i p(s,i)$
\begin{equation}
\frac{d}{dt}\vec{p}_s(t)=R\vec{p}_s(t)
\end{equation}
where $R$ is the associated rate matrix and $\vec{p}_s(t)=\left[p_s(1\da),p_s(1\ua),p_s(2\da),p_s(2\ua)\right]^{T}$. We want to include spin-conserving decay  $\Gamma_\da$ and $\Gamma_\ua$ from 2 into 1 and spin mixing $\gamma$ in both channels 1 and 2, so that
\begin{equation}
R=
\begin{pmatrix}
-\gamma & \gamma & \Gamma_{\da} &0 \\
\gamma & -\gamma & 0 & \Gamma_{\ua}\\
0 & 0 & - \Gamma_{\da} -\gamma & \gamma \\
0 & 0 & \gamma & -\Gamma_{\ua}-\gamma
\end{pmatrix}.
\end{equation}
The fit with a rate equation is only reliable with a normalized distribution. As evidenced by Fig.\ref{fig:Fits1}, this is not the case for this data and a strategy for normalization is required.

We considered different procedures to obtain a normalized $\vec{p}_s(t)$. Our first intention was to assume independent distributions $p(s,d)\simeq p_s(s)p_d(d)$, such that $p_s(s)\simeq p(s,1)/p_d(1)$. Nevertheless, it seems dynamics in the $s$ subspace increase the likelihood of $d=0$, so that distributions are somewhat correlated. Since $p(1\da)$ is largely static in Fig.\ref{fig:Fits1}(a), we only renormalize $p(1\ua)$ and $p(2\ua)$. We use full normalization for Fig.\ref{fig:Fits1}(b).

\begin{figure}
   \includegraphics[width=.7\columnwidth]{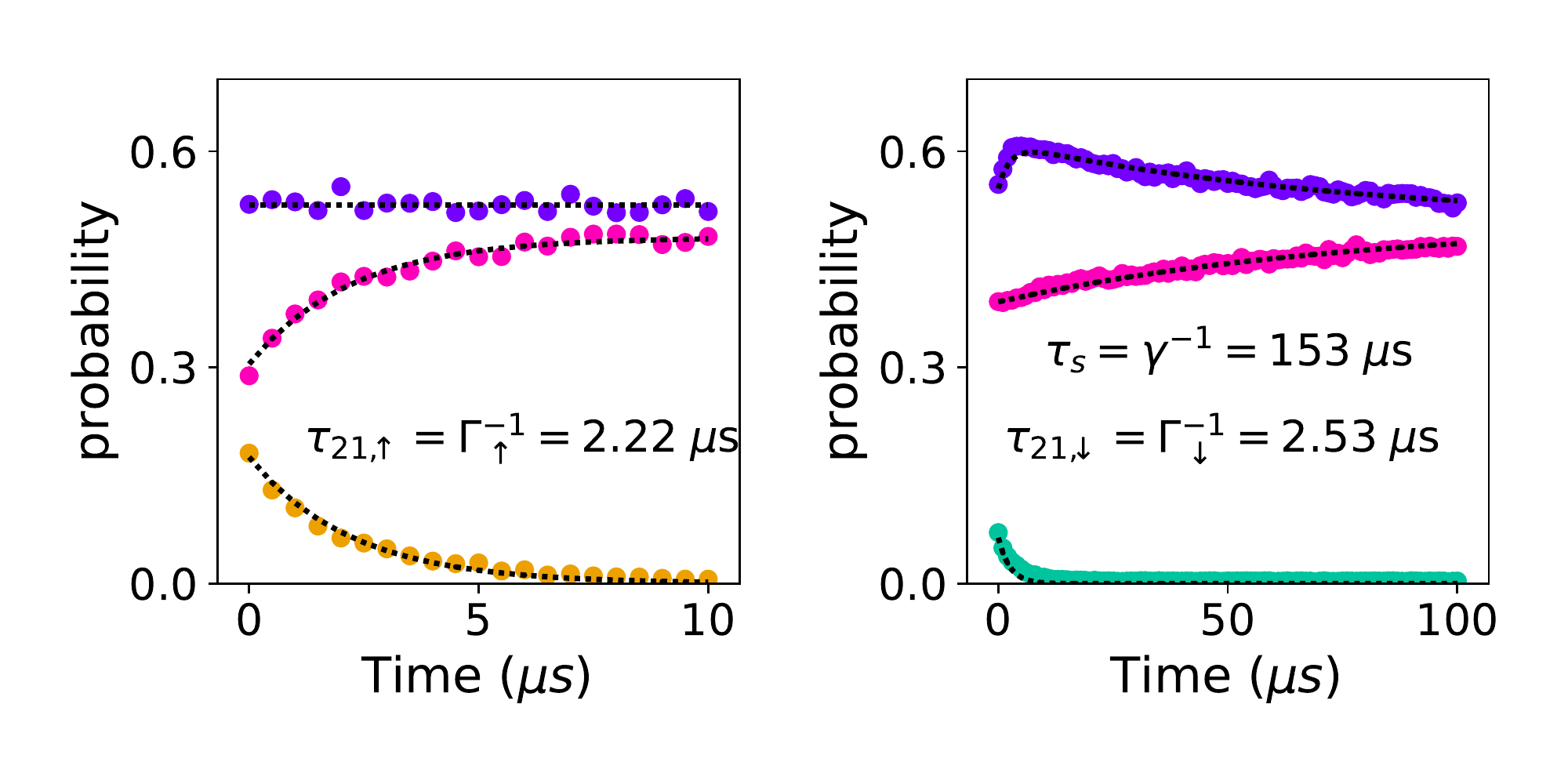}
\caption{Fits of the datapoints with the rate equation method described in the text.}
\label{fig:Fits2}
\end{figure}

We obtain values for $\gamma$, $\Gamma_{\da}$ and $\Gamma_{\ua}$, reproduced in Fig.\ref{fig:Fits2}. Whereas it seems confirmed that the mode 2 lifetimes are in the range of $2-3\mu$s, the spin lifetime produces a much larger value than the ones obtained in \cite{sHays2019}. We could refine our rate matrix to allow for non-full spin mixing, or discuss other normalization procedures. Nevertheless, all attempts at refitting the data have provided lifetimes well above $40\mu$s. It seems confirmed that this data indicates a longer spin lifetime than what is measured in Fig.3 of \cite{sHays2019}. In addition, the similarity of the values $\Gamma_{\da}$ and $\Gamma_{\ua}$ justifies the consideration of a spin independent mode 2 decay rate $\Gamma$ for simplicity.

\section{Rate Equation for Polarization}
We can take advantage of the faster dissipation rate of the mode levels in order to polarize the pseudospin degree of freedom by means of the application of a constant gate AC driving. Concepts of laser cooling can be useful for the optimization of this initial pulse.
In order to do so, we recognize that our polarization mechanism is analogous to a sideband cooling technique  in the context of laser cooling of trapped ions and follow derivations similar to those presented in \cite{sCirac1992}. An interaction picture with respect to $\w_g\tau_z/2$ and a RWA delivers the effective Hamiltonian
\be
H'=\frac{\Delta}{2} \tau_z + \frac{\W_g}{2}\tau_x + \boldsymbol{\tau}\cdot\boldsymbol{\sigma} + \frac{\nu}{2}\sigma_z
\ee
with the vectors $\boldsymbol{\tau}=\left(\tau_x,\tau_y,\tau_z\right)$ and $\boldsymbol{\sigma}=\left(\frac{\alpha\W_g}{2}\sigma_x,0,-\frac{g}{2}\sigma_z\right)$ and the detuning $\Delta=\w-\w_g$. The resonance choice of the main text corresponds to $\Delta=-\nu$, so that this is a protocol analogous to a sideband cooling scheme.

Now we derive a rate equation for the pseudospin populations. For this, we work in the interaction picture with respect to
\be
H_0=\frac{\Delta}{2} \tau_z + \frac{\W_g}{2}\tau_x - \frac{g}{2}\tau_z\sigma_z +  \frac{\nu}{2}\sigma_z,
\ee
and analyze the perturbation generated by $H_1=\frac{\alpha\W_g}{2}\tau_x\sigma_x$. We assume the adiabatic state of the system is
\be
\rho(t)=\sum_{s=\da,\ua}p_s(t)\prj{s}\rho_{s},
\ee
where $\rho_{s}$ is the steady state of the mode degrees of freedom given pseudospin polarization $s$. Pseudospin flip processes are only possible as second order effects encoded in the Born-Markov approximation
\be
\frac{d}{dt}\rho(t)=-\left[H_1(t),\int_0^t\left[H_1(t'),\rho(t)\right] dt' \right], \label{eq:BM}
\ee
so that the rates in equation
\be
\frac{d}{dt}
\begin{pmatrix}
p_\da\\
p_\ua
\end{pmatrix}
=
\begin{pmatrix}
-\gamma_\ua & \gamma_\da \\
\gamma_\ua & -\gamma_\da
\end{pmatrix}
\begin{pmatrix}
p_\da\\
p_\ua
\end{pmatrix},
\ee
correspond to
\bea
\gamma_s&=&2 Re\left[\int_0^\infty e^{\pm i\nu t}C_s (t)dt\right],\\
C_s(t)&=&Tr\left\{\tau_x\bra{s}U \ket{s}\tau_x \bra{\bar s}U^\dagger \ket{\bar s}\rho_{\bar s}\right\},
\label{eq:CF}
\eea
where $U=e^{-iH_0t}$. Note the lack of unitarity of the evolution and $ \bra{\bar s}U\ket{s}=0$. We approximate Eq.(\ref{eq:CF}) by means of the regression theorem and an associated modified master equation for the mode degrees of freedom alone represented by the Bloch equations
\be
\frac{d}{dt}
\langle
\boldsymbol{\tau}
\rangle^T
=
\begin{pmatrix}
-\frac{\Gamma}{2}-\gamma_d+b&\frac{\Delta}{2}+a&0\\
-\frac{\Delta}{2}-a&-\frac{\Gamma}{2}-\gamma_d+b&- \frac{\W_g}{2}\\
0& \frac{\W_g}{2}&-\Gamma
\end{pmatrix}
\langle
\boldsymbol{\tau}
\rangle^T
-
\begin{pmatrix}
0\\
0\\
\Gamma
\end{pmatrix}
\label{eq:BE}
\ee
The steady state $\rho_{\da,\ua}$ is computed with $b=0$ and $a=\pm\frac{g}{2}$ respectively, whereas the regression theorem for each correlation function $C_{\da,\ua}$ assumes Bloch equations with $a=0$ and $b=\mp i\frac{g}{2}$ respectively, which account for the lack of unitarity of the evolution. With this, it is possible to compute the real part of the half-sided Fourier transform of the correlation function Eq.(\ref{eq:CF}).

If we add the spin mixing processes, the dynamics of the populations are finally governed by the rate equation
\be
\frac{d}{dt}
\begin{pmatrix}
p_\da\\
p_\ua
\end{pmatrix}
=
\begin{pmatrix}
-\gamma_\ua - \gamma & \gamma_\da + \gamma \\
\gamma_\ua + \gamma & -\gamma_\da - \gamma
\end{pmatrix}
\begin{pmatrix}
p_\da\\
p_\ua
\end{pmatrix}.
\ee

\end{document}